\documentclass{cimento}
\usepackage{amsmath}
\usepackage{amsthm}
\usepackage{amssymb}
\usepackage[italian]{babel}
\usepackage[latin1]{inputenc}
\usepackage{epigraph}
\usepackage{expdlist}
\usepackage{epsfig}

\usepackage{graphicx}  

\title{The Mu2e experiment at Fermilab}
\author{R.~Donghia\from{ins:x} on behalf of the Mu2e Calorimeter group}
\instlist{\inst{ins:x} INFN-National Laboratory of Frascati  and Roma Tre University (Rome, Italy)}

\PACSes{
\PACSit{29.90}{+r}}

\begin{document}

\maketitle

\begin{abstract}
The Mu2e experiment searches for the neutrino-less muon to electron conversion in the field of a nucleus, which is a charged lepton flavor violating process. The goal of the experiment is to reach a single event sensitivity of $2.8 \times 10^{-17}$, setting an upper limit on the muon conversion rate at $6.7 \times 10^{-17}$. This corresponds to a four order of magnitude improvement with respect to the existing limits. 
\end{abstract}

\section{Charged Lepton Flavor Violation and Mu2e Physics}
The Mu2e experiment at Fermilab will search for the charged lepton flavor violating (CLFV) process of muon to electron conversion in an aluminum nucleus field, $\mu + N(Z, A) \rightarrow e + N(Z, A)$ \cite{ref:tdr}. No CLFV interactions have been observed experimentally yet. 

Lepton flavor violation for neutral leptons, proved by several experiments of the last decades, implies that also CLFV is possible even in the Standard Model (SM), but with a rate not-detectable by current experiments.
Indeed, in the SM, even considering diagrams with neutrinos oscillation, the expected rate is negligible ($ \sim~10^{-54}$). So, observation of these processes should be a crucial evidence of new physics beyond the SM. 

The current best experimental limit on $\mu-e$ conversion has been set by SINDRUM~II experiment \cite{SINDRUM}. Mu2e intends to improve its result by four orders of magnitude, constraining the ratio, $R_{\mu e}$, between the conversion rate and the number of muon captures by Al nucleus:
\begin{equation*}
R_{\mu e} = \frac{\mu^- \thinspace N(Z,A) \rightarrow e^- \thinspace N(Z,A)}{\mu^- \thinspace N(Z,A) \rightarrow\nu_{\mu} \thinspace N(Z-1,A)} < 6 \times 10^{-17} \thinspace , \thinspace \rm{at \thinspace 90\% \thinspace C.L.}
\end{equation*}
with a single event sensitivity of $2.7 \times 10^{-17}$
\section{The Mu2e apparatus}
The signature of this neutrinoless conversion process is a monoenergetic electron, with an energy slightly lower than the muon rest mass, $\sim$~104.96 MeV. In order to achieve our goal on the $R_{\mu e}$ measurement, a very intense muon beam ($\sim$ $10^{10}$~Hz) has to stop on an aluminum target and a precise momentum analysis has to be performed. The beam must also have a pulsed structure to discriminate the prompt beam-induced background. 
\begin{figure}[h!]
\centering
\includegraphics[width=.82\textwidth]{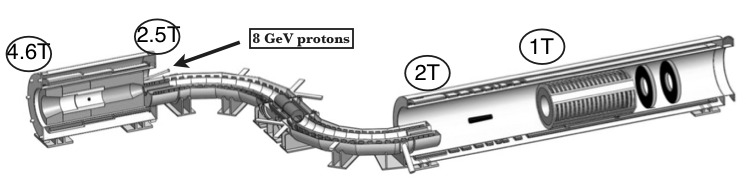}
\caption{Experimental layout of Mu2e. The Cosmic ray veto is not shown.}
\label{mu2e_layout}
\end{figure}

The Mu2e experiment~\cite{ref:tdr} is composed by three main
superconducting solenoid magnets (Fig.~\ref{mu2e_layout}). The
first is the production solenoid (PS), where an intense 8 GeV bunched proton beam strikes a tungsten target
producing mostly pions. Its graded magnetic field (from 4.6 T to 2.5 T) collects most of the produced charged particles and moves them through the second magnet, the transport solenoid (TS). The TS transfers and selects negative low-momentum muons towards the last detector solenoid (DS), trough a system of collimators. Its "S" shape prevents that neutral particles produced in the PS reach the last solenoid.
The DS houses the muon stopping Al targets and the
detection system devoted to identify and analyze the conversion
electrons (CEs). The Al targets reside in a graded field region (from 2~T to a 1~T), in this way electrons emitted upstream are reflected downstream through the detectors, whose acceptance is optimizes to collect $\sim$100~MeV/c tracks. 
The detector consists of a tracker made by 18 stations of straw tubes, with a resolution of $\sim$ 120 keV/c for 100 MeV/c
momentum, followed by a pure CsI crystal electromagnetic calorimeter, which allows particle identification and rejection of background. In particular, energy (with a resolution around 5\%, at 100 MeV) and timing (with a resolution better than 500 ps) measurements from the calorimeter provide critical information for efficient separation of electrons and muons in the detector.

A Cosmic Ray Veto system surrounds the DS on three sides ( the
ground is not covered) and extends up to the midpoint of the TS.
It is designed to reduce the number of expected
cosmic induced background events to 0.05 events during the entire
running period (three years).

\acknowledgments
This work was supported by the EU Horizon 2020 Research and Innovation Programme under the Marie Sklodowska-Curie Grant Agreement No. 690835.

\end{document}